\documentclass[12pt,preprint]{aastex}

\shorttitle{DIBs toward HD 62542}
\shortauthors{Snow et al.}

\received{2001 November 20}
\begin{document}

\title{Unusually Weak Diffuse Interstellar Bands toward HD 62542}

\author{Theodore P. Snow\altaffilmark{1},
Daniel E. Welty\altaffilmark{2,3},
Julie Thorburn\altaffilmark{4},
L. M. Hobbs\altaffilmark{4},
Benjamin J. McCall\altaffilmark{2,5,6},
Paule Sonnentrucker\altaffilmark{7}, and
Donald G. York\altaffilmark{2,5}}

\altaffiltext{1}{Center for Astrophysics and Space Astronomy, University of
Colorado, Boulder, CO 80309-0389;
tsnow@casa.colorado.edu}

\altaffiltext{2}{University of Chicago, Astronomy and Astrophysics 
Center, 5640 S. Ellis Ave., Chicago, IL  60637;
welty@oddjob.uchicago.edu, don@oddjob.uchicago.edu}

\altaffiltext{3}{visiting observer, Anglo-Australian Observatory}

\altaffiltext{4}{University of Chicago, Yerkes Observatory, Williams 
Bay, WI  53191;
thorburn@yerkes.uchicago.edu, hobbs@yerkes.uchicago.edu}

\altaffiltext{5}{also, Enrico Fermi Institute}

\altaffiltext{6}{current address: University of California, 601 
Campbell Hall, Berkeley, CA 94720;
bjmccall@astro.berkeley.edu}

\altaffiltext{7}{Johns Hopkins University, Department of Physics and 
Astronomy, 3400 N. Charles St., Baltimore, MD 21218;
sonnentr@pha.jhu.edu}

\begin{abstract}
As part of an extensive survey of diffuse interstellar bands (DIBs), we have
obtained optical spectra of the moderately reddened B5V star HD 62542, which
is known to have an unusual UV extinction curve of the type usually identified
with dark clouds.
The typically strongest of the commonly catalogued DIBs covered by the spectra
  --- those at 5780, 5797, 6270, 6284, and 6614\AA\ --- are essentially absent
  in this line of sight, in marked contrast with other lines of sight of similar
  reddening.
We compare the HD 62542 line of sight with others exhibiting a range 
of extinction properties and molecular abundances and interpret the weakness 
of the DIBs as an extreme case of deficient DIB formation in a dense cloud 
whose more diffuse outer layers have been stripped away.
We comment on the challenges these observations pose for identifying 
the carriers of the diffuse bands.
\end{abstract}

\section{Introduction}
\label{sec-intro}

Following the first reported detection of diffuse interstellar bands 
(DIBs) some 80 years ago (Heger 1922), several different approaches were 
made to the task of identifying the carriers (see Herbig 1995 or Snow 1995, 
2001 for recent reviews).
During the era of photographic spectroscopy, the principal approach 
was to examine correlations of the strengths of the DIBs with other interstellar 
quantities, but such studies rarely led to definitive conclusions 
(Snow, York, \& Welty 1977).
With the advent of electronic detectors in the 1970s, however, it 
became possible to measure the strengths and profiles of the DIBs more 
accurately.
Differences in profile shape among the DIBs and differences in 
relative DIB strength in different interstellar environments strongly suggest 
that the DIBs are due to a variety of carriers (Kre{\l}owski \& Walker 1987; 
Josafatsson \& Snow 1987; Kre{\l}owski \& Sneden 1993; Cami {\it et al.} 1997).
Some uncertainties remain, however, as to which of the DIBs are more 
closely related, because of differences both in the sightlines and DIBs 
considered and in how the DIBs are defined and measured in the various studies.

The breadth of the DIBs has often been ascribed to broadening by
predissocation, but this requires the DIB carriers to be continually
replenished at a rapid rate to maintain the necessary density of absorbers.
Douglas (1977) and Smith, Snow, \& York (1977) proposed a solution for this
dilemma: the DIB carriers could be large molecules undergoing internal
conversion, a process routinely seen in laboratory spectra.
Laboratory and theoretical studies of small and large ($>$10 atomic 
nuclei) molecules are now producing specific predictions of DIB candidate 
lines (e.g., Foing \& Ehrenfreund 1994 and Galazutdinov {\it et al.} 2000 
for C$_{60}$$^+$; Tulej {\it et al.} 1998 for C$_7^-$; McCarthy {\it et al.} 
1997, 2000 for carbon chains; Guthe {\it et al.} 2001 for l-C$_3$H$_2^-$; 
Kre{\l}owski {\it et al.} 2001 for naphthalene cations).
Ultra-high resolution spectra of several DIBs have revealed structure 
in the profiles suggestive of the rotational structure expected for large 
molecules (Sarre {\it et al.} 1995; Kerr {\it et al.} 1996; Ehrenfreund \& 
Foing 1996; Kre{\l}owski \& Schmidt 1997).
Astrophysical studies, together with the laboratory studies, thus 
have shown that large molecules may well be present in interstellar clouds 
and must be considered as possible candidates for some of the DIBs (Salama \& 
Allamandola 1992a,b; Snow {\it et al.} 1998; Sonnentrucker {\it et al.} 
1997, 1999).

We are currently assembling an extensive dataset of DIB information of much 
higher quality than has previously been available (Thorburn et al., in prep.). 
For most of the 60 stars in this survey, our optical spectra cover the range 
from 3700 to 10,500\AA, at a signal to noise ratio $\ga$1000 between 5780 and 
7000\AA.
For some Southern stars (including the one discussed in this paper), the spectra 
have less complete spectral coverage and lower S/N, however.
Our goal is to assemble a high-quality dataset --- including DIB strengths and 
many other interstellar parameters --- for a set of sightlines specifically 
selected to sample a wide variety of environments.
Our hope is that with sufficient data quality (allowing us to measure very weak 
DIBs and fine details of the DIB profiles), careful consideration of the 
continua near the DIBs, and more complete coverage of the other interstellar 
parameters, we may be able to identify definite dependences and relationships 
among the DIBs, between the DIBs and other interstellar species, and between the 
DIBs and specific physical or chemical factors characterizing the interstellar 
clouds.

In this paper, we report results for one particular line of sight with very 
unusual dust extinction properties and explore, by comparison with other lines 
of sight, the possible interpretations of the DIBs in this case.
In the next section, we summarize what is known about the star (HD 
62542) and its line of sight.  
In Section 3, we describe the observations and our results, while in 
Section 4 we discuss possible implications for understanding the DIBs.

\section{The Line of Sight toward HD 62542}
\label{sec-los}

The star HD 62542 lies behind material associated with the Gum Nebula 
(Cardelli \& Savage 1988; Cardelli {\it et al.} 1990; 
Whittet {\it et al.} 1993; Churchwell {\it et al.} 1996).
Whittet {\it et al.} estimated a spectroscopic distance of 405 pc, 
based on an adopted B3 V spectral type.
The distance estimate from {\it Hipparcos} --- 246 pc, with
a one-$\sigma$ uncertainty range from 212 to 294 pc --- is more 
consistent with the slightly later B5 V type adopted by Cardelli \& 
Savage, however.

Allowing for the uncertainties in spectral type, the color excess 
$E(B-V)$ = 0.33--0.37, the total visual extinction A$_{\rm V}$ =
1.07--1.20 mag, and the ratio of total to selective extinction  
R$_{\rm V}$ = 2.90--3.24 (Cardelli \& Savage 1988; Whittet {\it et al.} 1993; 
Rachford {\it et al.} 2002).  
The total hydrogen column density $N$(\ion{H}{1}) + 2$N$(H$_2$) is about 
2.2 $\times$ 10$^{21}$ cm$^{-2}$ (Rachford {\it et al.} 2002).

This star has one of the most extreme ultraviolet extinction curves known,
featuring a very broad, shortward-displaced 2175\AA\ bump and an
extremely steep far-UV rise --- similar to that of another extreme 
case, HD 29647 (Snow \& Seab 1980; Cardelli \& Savage 1988).  
Those extinction curve properties are often interpreted as indicating 
a preponderance of small dust grains.
Snow \& Seab (1980), Mathis \& Cardelli (1992), and Mathis (1994) 
have speculated that such extinction curves are produced by the addition 
of a thin mantle overlying the small carbonaceous grains thought to be 
responsible for the bump and the far-UV extinction.  
While a strong water ice feature has been observed 
toward the more heavily reddened HD 29647
(A$_{\rm V}$ = 3.6 mag; Goebel 1983), water ice mantles may not 
significantly affect the 2175\AA\ bump (Mathis 1994).
Given the much smaller value of A$_{\rm V}$ toward HD 62542,
we think detection of water ice in this sightline is unlikely ---
but would still recommend looking for it.

High resolution spectra of \ion{Na}{1} and \ion{K}{1} suggest that most of the
interstellar material in this line of sight is concentrated in a single cloud
at $v_{\odot}$ = 14 km s$^{-1}$; additional weaker components are 
seen in \ion{Na}{1} and \ion{Ca}{2} (D. Welty {\it et al.}, in prep.).
The column densities of CH, CN, and C$_2$ in the main cloud are quite 
high for the amount of reddening; CH$^+$ is not detected, however 
(Cardelli {\it et al.} 1990; Gredel, van Dishoeck, \& Black 1993).
Various diagnotics, interpreted via theoretical models, have 
suggested total hydrogen (H + H$_2$) densities $n_{\rm H}$ ranging from 
several hundred per cm$^3$ (Black \& van Dishoeck 1991) to 10,000 
cm$^{-3}$ (Cardelli {\it et al.}) in the main cloud.
The C$_2$ rotational level populations and the relative weakness of 
the mm-wave CN emission, however, suggest $T$ $\sim$ 50 K and $n_{\rm H}$ 
in the range 500--1000 cm$^{-3}$ (Gredel {\it et al.} 1991, 1993) --- 
also rather high for a cloud with such moderate extinction.
Spectra obtained with {\it FUSE} indicate a molecular hydrogen column 
density of 5.9 $\times$ 10$^{20}$cm$^{-2}$, with a corresponding hydrogen 
molecular fraction $f$(H$_2$) = 2$N$(H$_2$)/[$N$(\ion{H}{1}) + 2$N$(H$_2$)] = 
0.60, averaged over the line of sight (Rachford {\it et al.} 2002).
The molecular fraction is somewhat uncertain, however, owing to 
uncertainties in $N$(\ion{H}{1}).
If the density of the cloud is of order 500--1000 cm$^{-3}$, then the extent of
the cloud along the line of sight is only about 0.5--1.0 pc.
This cloud thus may be viewed as a small, dense knot within the 
filamentary structure associated with the Gum nebula.
Cardelli \& Savage (1988) and Cardelli {\it et al.} (1990) proposed 
that the more diffuse outer layers of the cloud may have been stripped by 
the passage of stellar winds and/or shocks within the Gum nebula.

\section{Observations and Results}
\label{sec-obs}

Table~\ref{tab:sdat} lists the stars compared in this paper, with 
spectral type, V magnitude, color excess, and a general characterization of 
the UV extinction curve.
Kappa Cas was chosen because it has nearly the same $E(B-V)$ and A$_{\rm V}$ 
as HD 62542.
It was the obvious difference between the DIBs toward $\kappa$ Cas 
and those toward HD 62542 that led us to write this paper --- the bands 
discussed in this paper are weaker in HD 62542 by factors of at least 4--8, despite the very similar $E(B-V)$ and A$_{\rm V}$.
Rho Oph D and $\theta^1$ Ori C were chosen for comparison because 
they belong to the class of stars with shallow far-UV extinction curves 
(Fitzpatrick \& Massa 1990) noted for having weak DIBs.
HD 183143 is included because it has frequently been cited in the 
literature as a standard for strong DIBs, for example in the compilations 
of Herbig (1975), Jenniskens \& D\'{e}sert (1994), and Tuairisg {\it et al.} 
(2000); HD 215733 is included as a lightly reddened star observed with the 
same instrument as HD 62542.

The spectra of HD 62542 and HD 215733 were obtained using the University
College London Echelle Spectrograph (UCLES; Walker \& Diego 1985) on the
Anglo-Australian Telescope in 1998 September.
Sections of 23 orders, giving incomplete coverage of the spectrum between 5075 
and 10,295\AA, were recorded on the MIT/LL2 CCD (2048$\times$4096; 
15$\mu$m pixels).
While the primary motivation for these observations was to obtain 
spectra of the interstellar \ion{Na}{1} D and \ion{K}{1} $\lambda$7698 lines, 
a number of DIBs also lie within the spectral coverage.
The spectra were extracted from the raw CCD images using standard procedures
within IRAF.
The spectral resolution is 5 km s$^{-1}$; for both stars, the S/N is of order
70 between about 5700 and 8800\AA.

The spectra of $\kappa$ Cas, $\theta^1$ Ori C, $\rho$ Oph D, and HD 183143 were
obtained at Apache Point Observatory (APO) with the Astrophysical 
Research Consortium (ARC) echelle spectrograph (ARCES).
The ARCES spectra are recorded at a resolution of 8 km s$^{-1}$, using a 1.6
arcsecond slit.
The CCD is a 2048$\times$2048 Site/Tektronix device with 24$\mu$m pixels.
Multiple spectra, with exposure times $\le$30 min, were obtained to achieve
S/N $\ga$ 1000 at 5780\AA.
For reddened stars, the S/N is greater than 100 throughout the entire 
range between 3600 and 10,000\AA.
The spectra were extracted using IRAF, with special settings to accommodate
the narrow (FWHM $\sim$ 3.3 pixels), closely spaced ($\sim$10 pixels, 
center to center) spectral orders and the aliasing caused by the fact that 
not all orders can be simultaneously lined up along the rows of CCD pixels.
All of the spectral orders passed by the system fit on the CCD.
Identical spectral regions from adjacent orders were weighted and 
co-added, after standard wavelength calibrations and one-dimensional flat 
field corrections were applied, to produce a blazeless spectrum covering the 
range 3600--10,000 \AA.
Details of the spectrograph and the routine observing procedures are discussed
by York, Hildebrand, Hobbs {\it et al.} (in prep.).

Figures~\ref{fig:sp57},~\ref{fig:sp62},~and~\ref{fig:sp66} show regions from
5725--5815\AA, 6200--6300\AA, and 6570--6670\AA. 
These wavelength regions are the only ones included in the UCLES spectra that 
contain strong, well-studied DIBs.
The top two spectra in each figure are noiseless atlas spectra (model 
representations of a flat continuum and Gaussian components designed to yield 
features comparable in strength to those seen toward HD 183143) from 
Jenniskens \& D\'{e}sert (1994; hereafter JD94) and 
Tuairisg {\it et al.} (2000; hereafter T00).
The atlases represent averages over several stars, with lines scaled 
in strength by the ratio of $E(B-V)$ in HD 183143 to that for each of the 
other stars.
The stellar spectra have been binned to a resolution of 8 km s$^{-1}$, 
shifted so that the interstellar \ion{K}{1} lines are at 0 km s$^{-1}$, 
and plotted in order of increasing A$_{\rm V}$ (top to bottom).
The dots on either side of the DIBs in the spectra of HD 183143 
denote the end points of the linear continua used for measuring equivalent 
widths.
The sharp features that occur in all stellar spectra at nearly the 
same wavelengths are telluric lines.
While there are many weak DIBs seen in common between JD94, T00, and 
our spectrum of HD 183143, there are also features that do not match up.
Some of these features are stellar lines (e.g., at 6579 and 6582\AA) 
which, of course, do not appear in the atlases; others are weak DIBs that are 
not well represented in the two atlases.
In addition, since the relative strengths of the DIBs can vary from 
star to star (Herbig 1995, and references therein), the averaged atlas
spectra do not necessarily match the actual spectrum of HD 183143 itself.
A more complete list of DIBs, which resolves discrepancies between 
previous atlases and our new spectra, is in preparation (York {\it et al.}, 
in prep.).

The DIB at 6284\AA\ deserves special mention, since it is partially
blended with a telluric band of O$_2$, and has long been recognized as a
difficult DIB to measure cleanly.  
In our data, however, the spectral resolving power is sufficient to clearly 
separate the narrow O$_2$ band features from the much broader DIB, and there was
thus no difficulty in excising the telluric absorption in order to measure the 
equivalent width of the diffuse band.  

The equivalent widths of five relatively strong, well-known DIBs for all six 
lines of sight are listed in Table~\ref{tab:isdat}, together with the ratio of 
total to selective extinction (R$_{\rm V}$), the total visual extinction 
(A$_{\rm V}$), and the column densities of \ion{H}{1}, H$_2$, \ion{K}{1}, 
and various molecular species.
Several footnotes to the table list measurements of the strengths of those DIBs 
from selected recent references, for comparison; a more thorough comparison of 
our new DIB measurements with previous values will be given by York {\it et al.} 
(in prep.).

\section{Discussion}
\label{sec-disc}

\begin{figure}
\plotone{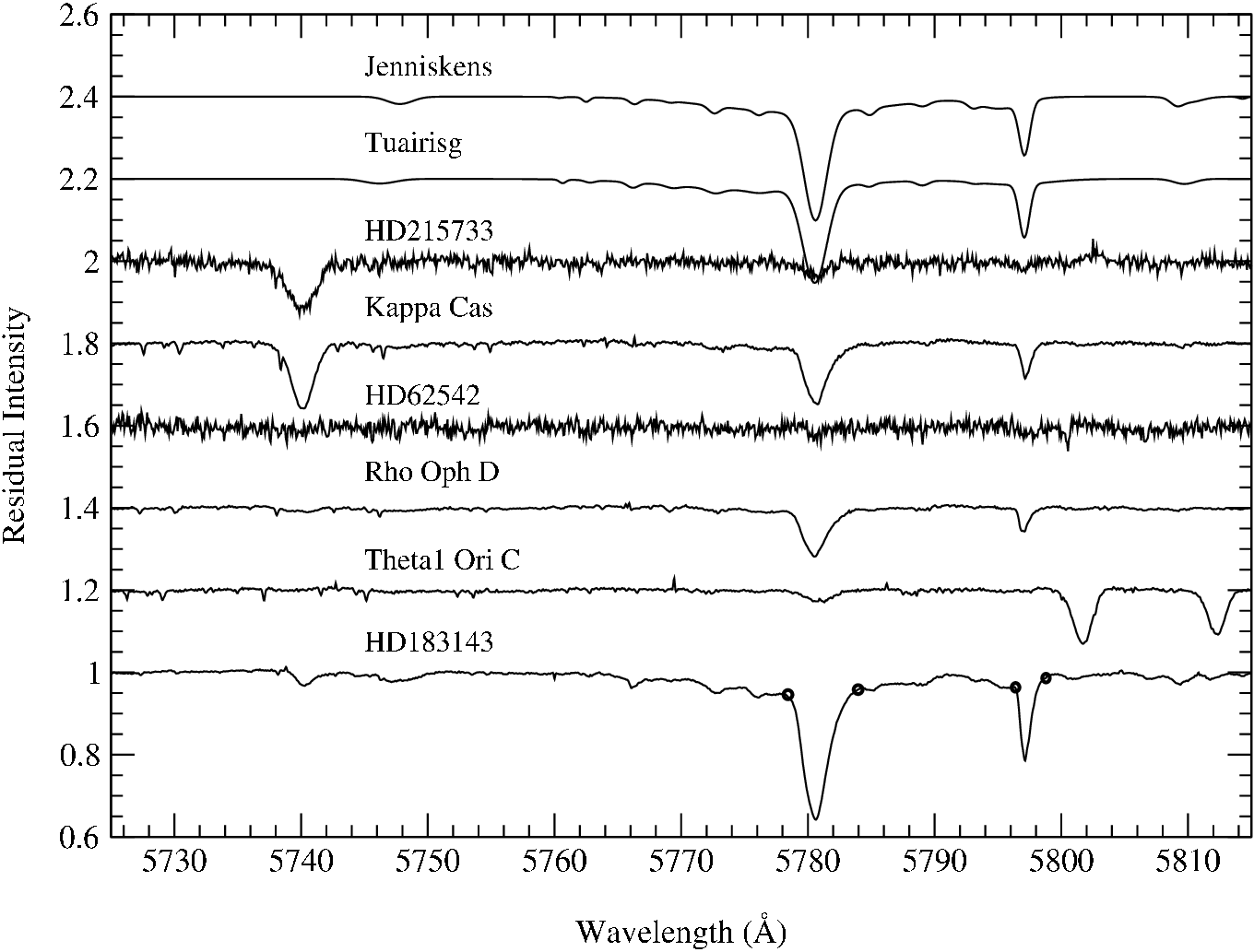}
\caption{The 5780\AA\ and 5797\AA\ DIBs toward six stars, arranged in 
order of increasing A$_{\rm V}$.
At the top are comparison spectra from the atlases of Jenniskens \& 
D\'{e}sert 1994 and Tuairisg et al. 2000.
Wavelengths have been shifted so that $v_{\rm LSR}$ = 0 km s$^{-1}$ 
for the K I $\lambda$7698 line --- so the
narrow telluric features seen in some spectra do not line up exactly.
The open circles on either side of the DIBs in HD 183143 mark the end 
points of the linear continua used for
determining equivalent widths.
HD 62542 clearly has much weaker DIBs than even the much less 
reddened HD 215733.}
\label{fig:sp57}
\end{figure}

\begin{figure}
\plotone{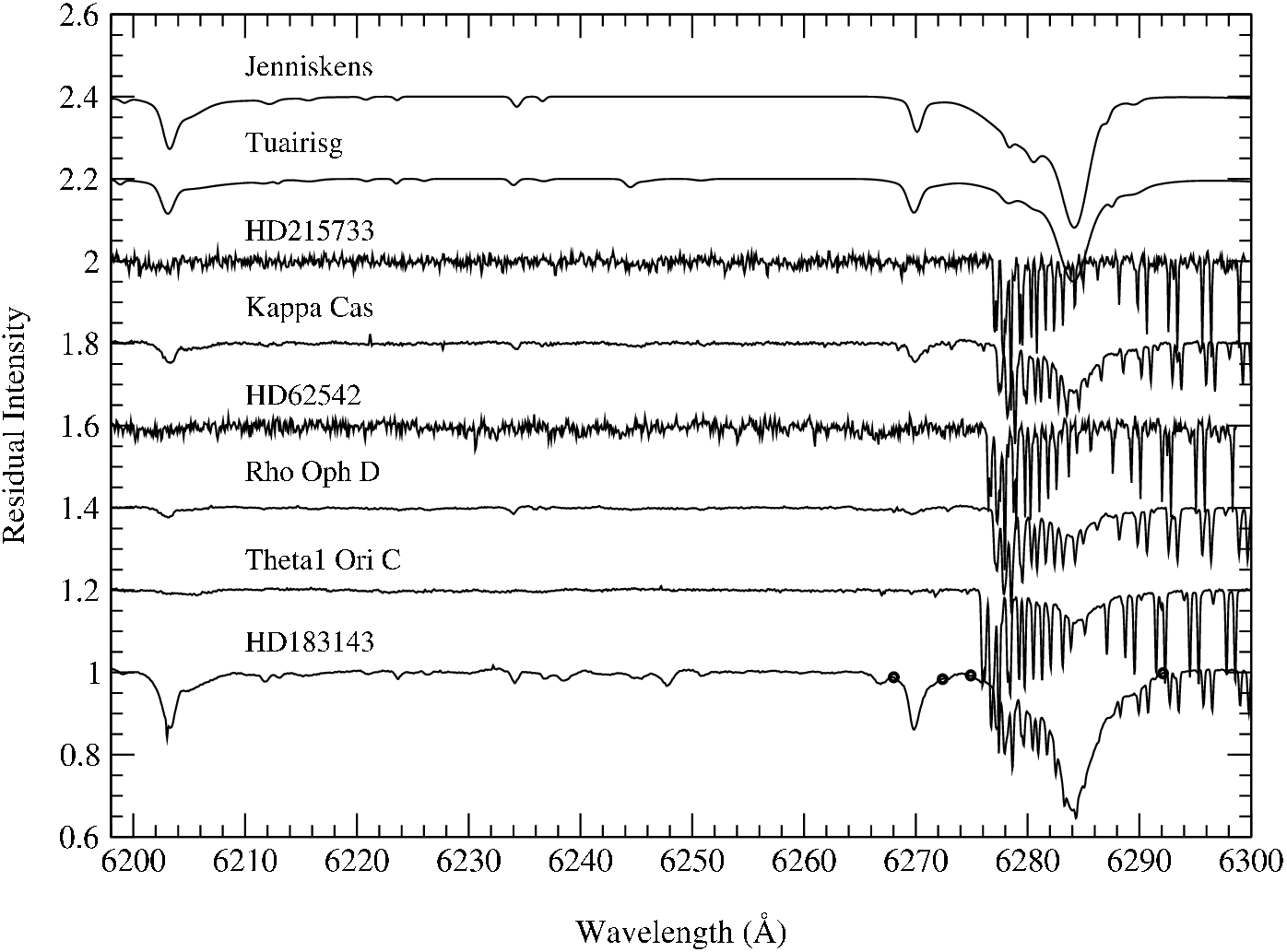}
\caption{The 6270\AA\ and 6284\AA\ DIBs, as in Fig. 1.
Note the strong telluric O$_2$ band absorption blended with the 6284\AA\ DIB (which does not, however, contribute to the tabulated equivalent widths of the DIB).}
\label{fig:sp62}
\end{figure}

\begin{figure}
\plotone{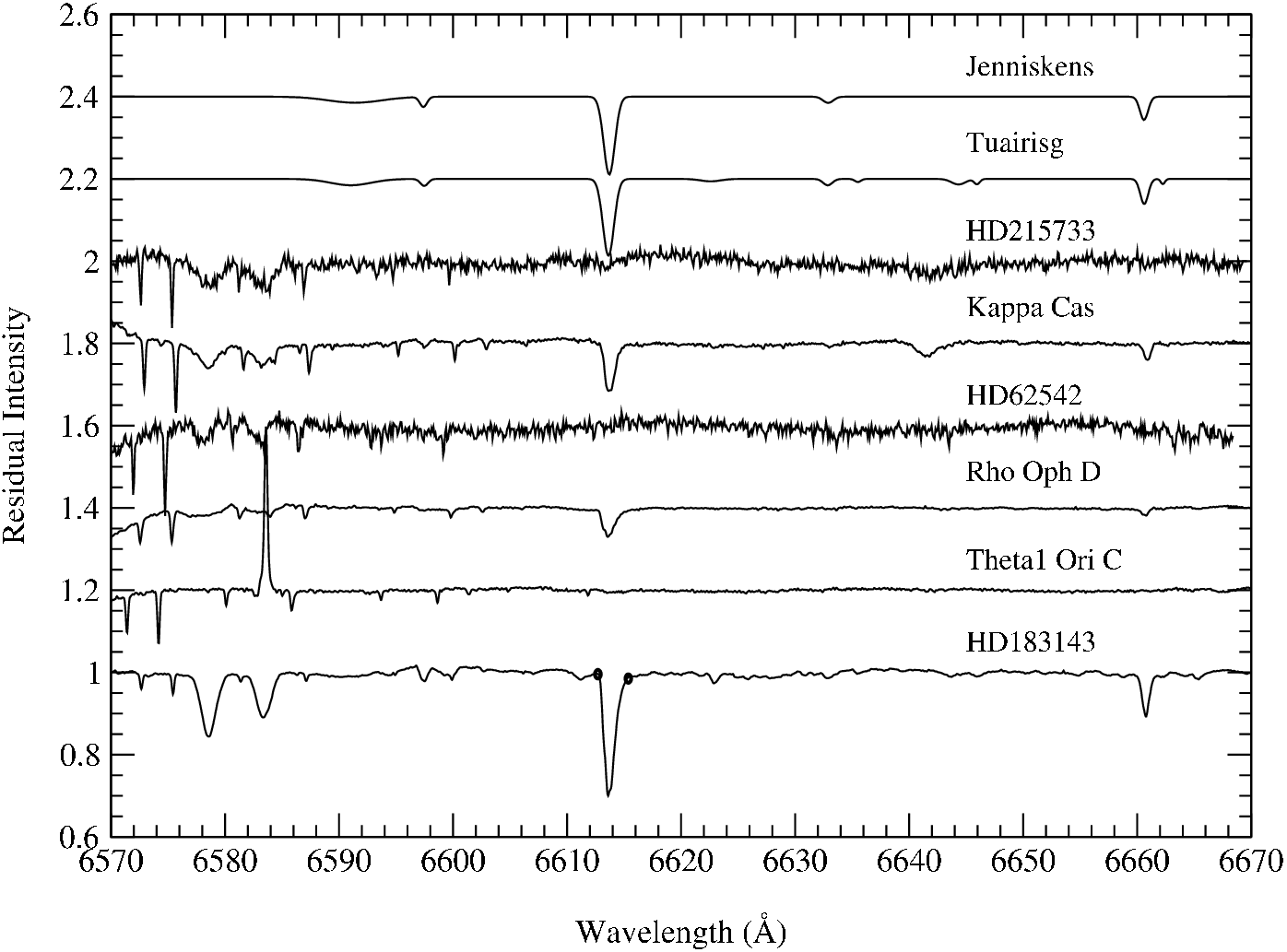}
\caption{The 6614\AA\ DIB, as in Fig. 1.}
\label{fig:sp66}
\end{figure}

It is apparent from Table~\ref{tab:isdat} and from 
Figures~\ref{fig:sp57}--~\ref{fig:sp66} that 
the strengths of the 5780, 5797, 6270, 6284, and 6614\AA\ DIBs do not increase 
monotonically with A$_{\rm V}$  --- at least for this small (and likely biased) 
sample of six lines of sight.  
As noted earlier, HD 2905 ($\kappa$ Cas) and HD 62542 have essentially identical A$_{\rm V}$, but the equivalent widths of all five DIBs are weaker toward 
HD 62542 by factors of at least 4--8.
None of those bands is detected toward HD 62542, given the S/N in our data, 
and we can list only upper limits in Table 2.\footnotemark
\footnotetext{ Several other, weaker DIBs identified in the literature 
(e.g., those at 6203 and 6660\AA) also are absent toward HD 62542, with 
upper limits similar to those for the other narrow DIBs.}
The nearly 70\% increase in A$_{\rm V}$ between HD 62542 and $\rho$ Oph D is 
accompanied by increases in the strengths of all five DIBs by factors of at 
least 2--6.
The bands toward $\rho$ Oph D are about half as strong as those 
toward $\kappa$ Cas, even though A$_{\rm V}$ is higher toward $\rho$ Oph D 
by a factor of almost 1.7.
The DIBs toward $\theta^1$ Ori C are even weaker (except 6284), even 
though A$_{\rm V}$ is slightly higher than toward $\rho$ Oph D.
On the other hand, A$_{\rm V}$ is greater by a factor of almost 3.9 toward 
HD 183143 than toward $\kappa$ Cas, and the five DIBs are stronger by factors 
of 2.5--3.4 --- more consistent with a nearly linear relationship.

\begin{figure}
\epsscale{1.0}
\plottwo{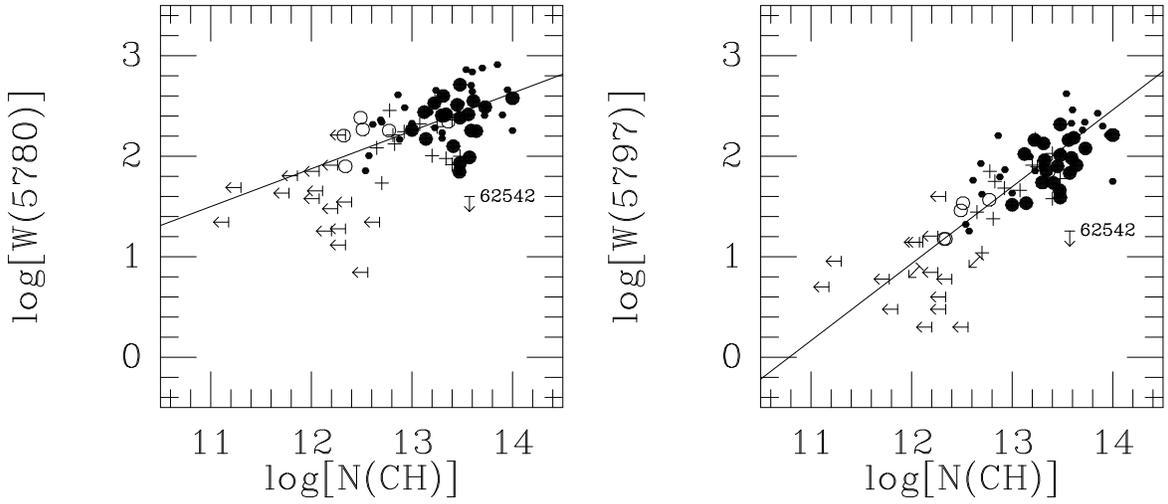}{f4b.eps}
\caption{The strengths of the DIBs at 5780 and 5797\AA\ versus the column density of CH.
Plus signs denote sightlines where $N$(CH) was derived from high-resolution spectra (FWHM $\la$ 2 km s$^{-1}$), while small filled circles have $N$(CH) from lower resolution spectra.
The larger filled circles represent translucent sightlines; open circles are for sightlines in the Sco-Oph region (see Welty \& Hobbs 2001).
The 5780\AA\ DIB is at best weakly correlated with $N$(CH), with slope = 0.38; the 5797\AA\ DIB shows a tighter correlation, with slope = 0.76.
Both DIBs are much weaker toward HD 62542 than for other sightlines with comparable $N$(CH).}
\label{fig:corr}
\end{figure}

Kre{\l}owski {\it et al.} (1999) have compared the equivalent widths of the 
DIBs at 5780 and 5797\AA\ with those of the strongest lines from the diatomic 
molecules CH (4300\AA) and CH$^+$ (4232\AA).
In Figure~\ref{fig:corr}, we plot the strengths (log W$_{\lambda}$) of these two 
DIBs versus the column density of CH.
The DIB strengths are primarily from our survey, but with some values from 
Josafatsson \& Snow (1987), Herbig (1993), and Kre{\l}owski {\it et al.} (1999);
the values for $N$(CH) are from various references (see the appendix to Welty \& 
Hobbs 2001), from fits to new high-resolution spectra of translucent sightlines 
(Welty {\it et al.}, in prep), and from our survey.
There is at best a weak correlation between the strength of the 5780\AA\ DIB 
and $N$(CH); the relationship between the strength of the 5797\AA\ DIB 
and $N$(CH) is somewhat tighter, with steeper slope.
Toward HD 62542, however, both DIBs are clearly much weaker than for other 
sightlines with comparable $N$(CH).
The DIBs toward HD 62542 are also anomalously weak when compared with the column 
densities of C$_2$ and CN, but such comparisons cannot be made with $N$(CH$^+$), 
as only an upper limit is available.

In view of the similarities in UV extinction toward HD 62542 and HD 29647, 
comparisons of other interstellar properties for these two lines of sight 
are of interest.
It is difficult to assess the atomic and molecular abundances toward 
HD 29647, due to its later spectral type (B8 V), its confusing spectrum 
(Hg-Mn), the near-perfect velocity match between the star and the main 
interstellar cloud in the line of sight (Adelman {\it et al.}2001), and 
the star's very low $v$sin$i$, which makes stellar lines appear nearly as 
narrow as interstellar lines in most currently available spectra (e.g. 
Crutcher 1985).
There is insufficient far-UV flux for {\it FUSE} observations, and it 
can be difficult to interpret, or even to identify unambiguously, some of 
the interstellar lines.
Owing to the breadth of the DIBs, however, it is possible to 
determine the DIB strengths toward this star.
It was previously established that the 5780 and 5797\AA\ DIBs toward 
HD 29647 are unusually weak, but might be present with very narrow 
profiles (Snow \& Seab 1991).
More comprehensive data to be presented by Snow {\it et al.} (in prep.) 
cover all the DIBs in Figs.~\ref{fig:sp57}--\ref{fig:sp66} here, and yield 
the following results:  
the 5780 and 5797\AA\ DIBs are present, with 5780 strongly dominant over 5797; 
the 6284\AA\ band is weakly present, with normal width; and the 6614\AA\ band 
is present but very weak.
The DIBs toward both stars thus are markedly weak relative to extinction, 
when compared to the DIBs in most diffuse cloud lines of sight reported in 
the literature (note that A$_{\rm V}$ = 3.6 mag for HD 29647).

There is longstanding evidence in the literature that the generically 
strongest DIBs tend to be relatively weak in dense clouds (e.g., 
Wampler 1966; Snow \& Cohen 1974; Adamson, Whittet, \& Duley 1991; 
Adamson {\it et al.} 1991; Snow 1995).
The current results for HD 62542 (and for HD 29647) appear to support 
this trend, in the extreme.
The DIBs are also known to be weak, relative to extinction, in
environments with intense radiation fields (Jenniskens, 
Ehrenfreund, \& Foing 1994; Snow {\it et al.} 1995).
This combination of factors --- DIB weakness in dense clouds {\it and} 
in regions of high UV flux --- leads to a picture in which the DIBs form 
and survive most readily either in ``ordinary'' diffuse clouds or in the 
diffuse outer layers of dense clouds (the ``skin effect'' discussed by 
Adamson {\it et al.} 1991).
The weakness of the DIBs toward HD 62542 and the suggestion of 
Cardelli {\it et al.} (1990) that the more diffuse outer layers of the 
dense cloud toward HD 62542 have been stripped off are consistent with 
this scenario.

Of the sightlines considered in this paper, the three with ``normal'' 
DIB strengths per unit $E(B-V)$ (HD 215733, $\kappa$ Cas, HD 183143) all 
have a number of interstellar components --- of roughly comparable strength 
and spread out in velocity --- in \ion{C}{1}, \ion{Na}{1}, 
and/or \ion{K}{1} (Fitzpatrick \& Spitzer 1997; Welty \& Hobbs 2001; 
McCall {\it et al.} 2002).
Toward $\kappa$ Cas and HD 183143, $N$(CN)/$N$(CH) $<$ 0.05 and 
$N$(CH$^+$)/$N$(CH) $\ga$ 1.0 (Table~\ref{tab:isdat}) --- both suggestive 
of relatively low densities in an ensemble of ``normal'' diffuse clouds.
The sightlines with weak DIBs (HD 62542, $\rho$ Oph D, $\theta^1$ 
Ori C), however, are characterized by somewhat different structures 
and/or atomic and molecular abundances.
While there are a number of \ion{Na}{1} components toward $\theta^1$ Ori C, 
the overall $N$(\ion{Na}{1}) is quite low, relative to $N$(H) (Welty \& 
Hobbs 2001), and CH, CN, and CH$^+$ have not been detected --- all 
suggesting that those components correspond to diffuse
clouds bathed in a strong ambient radiation field.
The available high resolution \ion{Na}{1} and \ion{K}{1} spectra of 
HD 62542 and $\rho$ Oph D indicate, however, that those lines of sight are 
each dominated by a single strong component [or perhaps several narrow, 
closely blended components for $\rho$ Oph D, as seen
in higher resolution spectra of $\rho$ Oph A (Welty \& Hobbs 2001)].
The molecular column densities toward HD 62542 imply the presence of much 
denser gas:  $N$(CN)/$N$(CH) $\sim$ 1.0 and $N$(CH$^+$)/$N$(CH) $<$0.02;
$\rho$ Oph D represents an intermediate case, with $N$(CN)/$N$(CH) $\sim$ 0.1 
and $N$(CH$^+$)/$N$(CH) $\sim$ 0.3.

The carriers of the DIBs considered here thus seem to require a 
relatively low density and/or (at most) a moderately intense radiation 
field (Cami {\it et al.}, 1997; Sonnentrucker {\it et al.} 1997).
Correlations between the strengths of the 5780 and 5797\AA\ DIBs, 
$N$(H), $N$(\ion{Na}{1}), and $N$(\ion{K}{1}) suggest that the DIB 
carriers generally behave like dominant ions --- though the weakness of 
those DIBs toward stars in Sco-Oph and in the Orion Trapezium
region is reminiscent of the weakness of the trace neutral species 
there (Herbig 1993; Welty \& Hobbs 2001).
There also are indications that the carrier of the 5797\AA\ DIB may 
be more readily destroyed (ionized?) than that of the 5780\AA\ DIB where 
the radiation field is stronger (Sonnentrucker {\it et al.} 1997; 
Kre{\l}owski, Galazutdinov, \& Musaev 1998; Welty \& Hobbs 2001).
Such patterns are consistent with suggestions that the DIB carriers 
are molecular ions --- which are destroyed (through electron
attachment) in dense environments (e.g., toward HD 62542) or which 
are not formed or maintained in clouds with weak radiation fields (e.g., 
Leger \& d'Hendecourt 1985; Crawford, Tielens, \& Allamandola 1985; 
van der Zwet \& Allamandola 1985; Salama {\it et al.} 1996, 1999; see 
LePage, Snow, \& Bierbaum 2001 for a recent overview and model).
No specific molecule or family of molecules can be firmly identified 
at the present time, however.
The DIB behavior argues strongly against molecular anions as the DIB 
carriers --- at least for the stronger, more well-studied DIBs --- 
consistent with the recent rejection of the C$_7^-$ hypothesis
(McCall {\it et al.} 2001; Galazutdinov {\it et al.} 2001; Sarre \& 
Kendall 2000).

While we are tempted to associate the weakness of the DIBs toward 
HD 62542 with the low UV field intensity and high molecular content of the 
main cloud at 14 km s$^{-1}$, we note that the hydrogen
molecular fraction is not extremely high.
The value $f$(H$_2$) = 0.60 (if verified once a more accurate \ion{H}{1} 
column density can be obtained) is comparable to, but not greater than, 
the values found for other diffuse or translucent
clouds such as the main clouds toward $\zeta$ Oph.
We hypothesize instead that the part of the main cloud toward HD 62542 
that is conducive to the formation and survival of DIBs is unusually 
thin, sandwiched between the dense cloud that dominates
the molecular absorption and the hot, low-density gas of the Gum 
nebula in which that cloud is immersed.
In this view, the more extended diffuse neutral region where the DIB 
carriers would normally reside is essentially absent in this line of sight.
The main cloud toward HD 29647, with somewhat stronger DIBs than 
those toward HD 62542, appears to be in a more quiescent region 
(Cardelli \& Savage 1988), and thus may have a more normal outer
structure.

\section{Summary/Conclusions}
\label{sec-summ}

A number of the diffuse interstellar bands (at 5780, 5797, 6270, 
6284, and 6614 \AA) commonly seen even in relatively lightly reddened 
lines of sight appear to be essentially absent toward the
moderately reddened star HD 62542, previously known to exhibit both 
strong interstellar molecular lines and an unusually steep far-UV 
extinction curve with a weak, shifted 2175\AA\ bump.
The line of sight appears to be dominated by a single small, very 
dense cloud, whose more diffuse outer layers may have been stripped by 
stellar winds and/or shocks within the Gum nebula.

Comparisons of those five moderately strong DIBs in six lines of 
sight indicate that the strengths of the DIBs are not well correlated with 
A$_{\rm V}$, but must depend on other (as yet undetermined)
properties of the interstellar clouds.
For this small sample, the bands are weaker, per unit A$_{\rm V}$, 
for sightlines with strong CN and weak CH$^+$, relative to CH (HD 29647, 
HD 62542), and are stronger for sightlines with
weak CN and strong CH$^+$ ($\kappa$ Cas, HD 183143) --- consistent 
with reduced abundances of the carriers of these DIBs in denser gas.
[We note that our larger sample suggests that different DIBs can show somewhat
different dependences on $N$(CH$^+$)/$N$(CN), however.]
Moreover, the DIBs are weaker when the far-UV extinction is either 
very steep, and the radiation fields not enhanced (HD 62542, 
HD 29647) {\it or} very shallow, with a strong ambient field 
($\theta^1$ Ori C); the other sightline with relatively weak DIBs 
($\rho$ Oph D) has moderately shallow far-UV extinction and (probably)
a somewhat enhanced radiation field.
Thus, too little {\it or} too much UV flux removes the DIB absorbers 
from the gas --- accounting qualitatively for the very large variations in 
DIB strengths with A$_{\rm V}$ seen in Table~\ref{tab:isdat}.
The results presented in this paper thus extend the similar 
conclusions of Cami {\it et al.} (1997) to higher densities and 
molecular abundances.
The absence of DIBs toward HD 62542, combined with similar behavior 
seen in other sightlines with similar interstellar conditions, suggests 
that molecular ions, or other species having a similar
response to ambient density and radiation, are the most likely 
carriers of the DIBs.

We are assembling a data base for 60 lines of sight --- with data on 
H$_2$, CH, CH$^+$, CN, C$_2$, A$_{\rm V}$, $f$(H$_2$), T$_{01}$(H$_2$), 
T$_{35}$(H$_2$); detailed component structures from high-resolution 
spectra of \ion{K}{1}, \ion{Na}{1}, CH, CH$^+$, and CN
(Welty \& Hobbs 2001; D. Welty {\it et al.}, in prep.); abundances of 
H$_3^+$ and CO (McCall {\it et al.} 2002); strengths of interstellar IR 
emission and absorption bands; and element abundances/depletions measured 
with {\it HST} and {\it FUSE} (Sonnentrucker {\it et al.}, in prep.) --- 
in order to search for correlations between the DIB strengths and other 
interstellar properties.

\acknowledgments

TPS acknowledges NASA grant NAG5-6758 and NASA contract 2430-60020, 
both to the University of Colorado.
DEW acknowledges support from NASA grant NAG5-3228 to the University 
of Chicago.
BJM acknowledges support from the Fannie and John Hertz Foundation and
the Miller Institute for Basic Research in Science.
PS acknowledges support from NASA contract NAS5-32985 to Johns 
Hopkins University.
This work is based in part on observations obtained with the Apache 
Point Observatory 3.5 m telescope, which is owned and operated by the 
Astrophysical Research Consortium.

\clearpage

\clearpage

\begin{deluxetable}{rllccl}
\tabletypesize{\scriptsize}
\tablecolumns{6}
\tablenum{1}
\tablecaption{Stellar Data\label{tab:sdat}}
\tablewidth{0pt}

\tablehead{
\multicolumn{1}{l}{HD}&
\multicolumn{1}{c}{Name}&
\multicolumn{1}{c}{Type\tablenotemark{a}}&
\multicolumn{1}{c}{$V$\tablenotemark{a}}&
\multicolumn{1}{c}{$E(B-V)$\tablenotemark{b}}&
\multicolumn{1}{l}{UV extinction\tablenotemark{c}}}
\startdata
  2905 & $\kappa$ Cas     & B1 Iae & 4.16 & 0.33 & average \\
 37022 & $\theta^1$ Ori C & O6     & 5.13 & 0.34 & very shallow far-UV \\
 62542 & \nodata          & B5 V   & 8.04 & 0.35 & very steep far-UV \\
147888 & $\rho$ Oph D     & B5 V   & 6.74 & 0.47 & shallow far-UV \\
183143 & \nodata          & B7 Ia  & 6.86 & 1.27 & slightly shallow far-UV\\
215733 & \nodata          & B1 II  & 7.32 & 0.10 & \nodata \\
\enddata

\tablenotetext{a}{Hoffleit 1982; Hoffleit {\it et al.} 1983; Cardelli \& Savage 1988; Albert 1983.}
\tablenotetext{b}{Using the intrinsic colors of Johnson 1963.}
\tablenotetext{c}{Cardelli \& Savage 1988; Fitzpatrick \& Massa 1990; 
Jenniskens \& Greenberg 1993.}

\end{deluxetable}

\begin{deluxetable}{lcccccc}
\tabletypesize{\scriptsize}
\tablecolumns{7}
\tablenum{2}
\tablecaption{Interstellar Data\tablenotemark{a}\label{tab:isdat}}
\tablewidth{0pt}

\tablehead{
\multicolumn{1}{c}{ }&
\multicolumn{1}{c}{ }&
\multicolumn{1}{c}{$\kappa$ Cas}&
\multicolumn{1}{c}{ }&
\multicolumn{1}{c}{$\rho$ Oph D}&
\multicolumn{1}{c}{$\theta^1$ Ori C}&
\multicolumn{1}{c}{ }\\
\multicolumn{1}{c}{Quantity}&
\multicolumn{1}{c}{HD 215733}&
\multicolumn{1}{c}{HD 2905\tablenotemark{b}}&
\multicolumn{1}{c}{HD 62542}&
\multicolumn{1}{c}{HD 147888}&
\multicolumn{1}{c}{HD 37022\tablenotemark{c}}&
\multicolumn{1}{c}{HD 183143\tablenotemark{d}}}
\startdata
A$_{\rm V}$ & 0.31 & 1.02 & 1.02 & 1.71 & 1.87 & 3.94 \\
R$_{\rm V}$ &(3.1) &(3.1) & 2.90 & 3.63 & 5.50 &(3.1) \\
& \\
log$N$(H I) (cm$^{-2}$)           & 20.70 & 21.20 & 20.93 &\nodata& 
     21.54 &\nodata \\
log$N$(H$_2$) (cm$^{-2}$)         &\nodata& 20.27 & 20.81 &\nodata& 
  $<$17.55&\nodata \\
log$N$(H$_{\rm tot}$) (cm$^{-2}$) &\nodata& 21.29 & 21.33 &\nodata& 
     21.54 &\nodata \\
log$f$(H$_2$)                     &\nodata&$-$0.72&$-$0.22&\nodata&
  $<-$3.69&\nodata \\
& \\
W(5780) (m\AA) & 76$\pm$15 & 304$\pm$10 & $<$40  & 254$\pm$5  &
     62$\pm$8   &  751$\pm$10 \\
W(5797) (m\AA) & 19$\pm$6  &  74$\pm$4  & $<$18  &  55$\pm$3  &
     $<$12      &  182$\pm$4  \\
W(6270) (m\AA) & $<$12     &  55$\pm$5  & $<$12  &  20$\pm$4  &
     $<$10      &  188$\pm$5  \\
W(6284) (m\AA) & $<$120    & 612$\pm$45 & $<$120 & 415$\pm$40 &
     460$\pm$60 & 1854$\pm$60 \\
W(6614) (m\AA) & $<$15     & 123$\pm$3  & $<$15  &  82$\pm$4  &
     10$\pm$3   &  334$\pm$6  \\
& \\
log$N$(CH) (cm$^{-2}$)     &\nodata&   12.93 &   13.58 &   13.30 &$<$11.86 &
     13.70 \\
log$N$(CN) (cm$^{-2}$)     &\nodata&$<$11.29 &   13.62 &   12.30 &$<$12.52 &
     12.34 \\
log$N$(C$_2$) (cm$^{-2}$)  &\nodata& \nodata &   13.90 &   13.34 &\nodata &
  $<$12.85 \\
log$N$(CH$^+$) (cm$^{-2}$) &\nodata&   13.21 &$<$11.82 &   12.81 &$<$12.00 &
     13.67 \\
& \\
log$N$(K I) (cm$^{-2}$)    & 11.08 &   11.71 &   11.90 &$>$11.98 & 10.48 &
     12.21 \\
\enddata

\tablenotetext{a}{H I and H$_2$ data are from Bohlin {\it et al.} 
1978; Savage {\it et al.} 1977; Diplas \& Savage 1994; and Rachford 
{\it et al.} 2002.
Molecular data are from Cardelli {\it et al.} 1990; Federman (priv. 
comm.); Gredel {\it et al.} 1991, 1993; Gredel 1999; Price {\it et al.} 2001; McCall {\it et al.} 2002; and Welty {\it et al.} (in prep.).
K I data are from Welty \& Hobbs 2001; Welty {\it et al.} (in prep.); 
and McCall {\it et al.} 2002.
Values of R$_V$ are from Rachford {\it et al.} 2002 and Cardelli {\it et al.} 1989; values in parentheses are based on the standard interstellar value.}
\tablenotetext{b}{Kre{\l}owski {\it et al.} 1999 list W(5780) = 280 m\AA, W(5797) = 70 m\AA\ (averages of four values each).}
\tablenotetext{c}{Herbig 1993 lists W(5780) = 64 m\AA, W(5797) = 3 m\AA; Jenniskens {\it et al.} 1994 list W(6270) = 14 m\AA, W(6284)= 480 m\AA.}
\tablenotetext{d}{Herbig 1995 lists W(5780) = 801 m\AA, W(5797) = 238 m\AA, W(6270) = 222 m\AA, W(6284) = 1945 m\AA, W(6614) = 358 m\AA.
Tuairisg {\it et al.} 2000 list W(5780) = 770 m\AA, W(5797) = 187 m\AA, W(6284) = 2003 m\AA\ (sum over several DIBs included in our limits), W(6614) = 320 m\AA.}

\end{deluxetable}

\end{document}